\documentclass[12pt]{iopart}
\usepackage{graphicx}
\begin{document}
\title{Model independent analysis of dark energy: Supernova fitting result}
\author{Yungui Gong}
\address{College of Electronic
Engineering, Chongqing University of Posts and Telecommunications,
Chongqing 400065, China} \ead{gongyg@cqupt.edu.cn}
\begin{abstract}
This paper uses the supernova data to explore the property of dark
energy by some model independent methods. We first Taylor expand
the scale factor $a(t)$ and the luminosity distance $d_{\rm L}$ to
the fifth order to find out that the deceleration parameter
$q_0<0$. This result just invokes the Robertson-Walker metric. So
the conclusion that the universe is expanding with acceleration is
more general. Then we discuss several different parametrizations
used in the literature. We also proposed two modified
parametrizations. We find that $\omega_{\rm DE0}$ is less than
$-1$ almost at $1\sigma$ level from all the parametrizations used
in this paper. We also find that the transition redshift from
deceleration phase to acceleration phase is $z_{\rm T}\sim 0.3$.
\end{abstract}
\maketitle

\section{Introduction}

The type Ia supernova (SN Ia) observation suggests that dark
energy contributes 2/3 to the critical density of the present
universe \cite{sp99,gpm98,agr98}. SN Ia observation also provides
the evidence of a decelerated universe in the recent past with the
transition redshift $z_{\rm T}\sim 0.5$ \cite{agr,mstagr,riess}.
The cosmic microwave background (CMB) observations favor a
spatially flat universe as predicted by inflationary models
\cite{pdb00,sh00}. There are many dark energy models proposed in
the literature. For a review of dark energy models, see, for
example, \cite{review} and \cite{padmanabhan03} and references
therein. However, the nature of dark energy is still unknown. It
is not practical to test every single dark energy model by using
the observational data. Therefore, a model independent probe of
dark energy is one of the best ways to study the nature of dark
energy.

The type Ia supernovae (SNe Ia) as standard candles are used to
measure the luminosity distance-redshift relationship $d_{\rm
L}(z)$. So we can model the luminosity distance $d_{\rm L}$ to
study the property of dark energy. Melchiorri etal. first found
that dark energy may be a phantom type by combining different
observational data to probe the behaviour of dark energy
\cite{melchiorri}. Huterer and Turner modelled the luminosity
distance by a simple power law $d_{\rm L}(z)=\sum_{i=1}^{N}c_iz^i$
\cite{turner}. Saini etal. used a more complicated function to
model the luminosity distance \cite{saini}. Another way to probe
the nature of dark energy is to parameterize the dark energy
equation of state parameter $\omega_{\rm DE}$. The simplest
parametrization is the constant equation of state model
$\omega_{\rm DE}={\rm constant}$. Several authors modelled
$\omega_{\rm DE}$ as $\omega_{\rm DE}=\sum_{i=0}^N \omega_i z^i$
\cite{jwaa,jwaa1,pastier}. Apparently this parametrization is not
good for high $z$. Recently, a stable parametrization $\omega_{\rm
DE}=\omega_0+\omega_a z/(1+z)$ was used in
\cite{polarski,linder,linder1,choudhury}. By fitting the model to
SN Ia data, we find that $\omega_0+\omega_a>0$, so this
parametrization is not good at high $z$ too. Jassal, Bagla and
Padmanabhan modified this parametrization as $\omega_{\rm
DE}=\omega_0+\omega_a z/(1+z)^2$ and the problem was solved
because $\omega_{\rm DE}=\omega_0$ at present and at high $z$
\cite{hkjbp}. More complicated functional forms for $\omega_{\rm
DE}(z)$ were also proposed in the literature
\cite{efstathiou,gefstathiou,pscejc,wetterich}. We can also model
the dark energy density itself. For example, a simple power law
expansion $\Omega_{\rm DE}=\sum_{i=0}^N A_i z^i$ was used to
investigate the nature of dark energy
\cite{alam,alam1,daly,daly1,gong,jonsson}. There are other
parametrizations, like the piecewise constant parametrization
\cite{wang,wang1,cardone,huterer}.

This paper is organized as follows. In section II, We first use a
Taylor expansion to expand the scale factor, then we fit the model
to the whole 157 gold sample of SNe Ia compiled by Riess etal. in
\cite{riess}. By expanding the scale factor, the fitting
parameters have physical meanings. In section III, we analyze the
dark energy parametrization proposed by Alam etal. \cite{alam}. In
section IV, we first study the parametrization $\omega_{\rm
DE}=\omega_0+\omega_a z/(1+z)$ and point out that this
parametrization is not good at high $z$. Then we study the
parametrization $\omega_{\rm DE}=\omega_0+\omega_a z/(1+z)^2$. In
section V, we first investigate the parametrization proposed by
Wetterich \cite{wetterich}, then we propose two modified
parametrizations. In section VI, we give some discussions.

\section{Model Independent Method}
In a homogeneous and isotropic universe, the
Friedmann-Robertson-Walker (FRW) space-time metric is
\begin{equation}
\label{rwmetric} ds^2=-dt^2+a^2(t)\left[{dr^2\over
1-k\,r^2}+r^2\,d\Omega\right].
\end{equation}
For a null geodesic, we have \begin{equation} \label{line}
\int_{t_1}^{t_0}{dt\over a(t)}=\int_0^{r_1}{dr\over
\sqrt{1-kr^2}}\equiv f(r_1), \end{equation} where
\begin{equation}
f(r_1)=\left\{\begin{array}{ll}
\sin^{-1}r_1,\ \ \ \ \ \ &k=1,\\
r_1,&k=0,\\
\sinh^{-1}r_1,&k=-1.
\end{array}\right.
\end{equation}
From Eq. (\ref{line}), we get the luminosity distance $d_{\rm
L}=a_0(1+z)r_1$ by Taylor expansion \cite{visser},
\begin{eqnarray}
\label{lumdis} H_0d_{\rm L}&=&z+{1\over 2}(1-q_0)z^2+{1\over
6}(q_0+3q_0^2-1-j_0-\Omega_k)z^3+ {1\over
24}(2-2q_0-\nonumber\\
&&15q_0^2-15q_0^3+5j_0+10q_0j_0+s_0+2\Omega_k+6\Omega_kq_0)z^4+O(z^5),
\end{eqnarray}
where the redshift $z$ is defined as $1+z=a_0/a(t)$, the subscript
0 means that a variable is evaluated at the present time, the
Hubble parameter $H(t)$, the deceleration parameter $q(t)$, the
jerk parameter $j(t)$ and the snap parameter $s(t)$ are defined as
\begin{equation} H(t)=\dot{a}/a={1\over a}{da\over dt},
\end{equation}
\begin{equation}
q(t)=-a^{-1}H^{-2}\ddot{a}=-{1\over aH^2}{d^2a\over
dt^2},
\end{equation}
\begin{equation}
j(t)={1\over aH^3}{d^3a\over
dt^3},
\end{equation}
\begin{equation}
s(t)={1\over aH^4}{d^4a\over
dt^4},
\end{equation}
and $\Omega_k=k/(a^2_0H^2_0)$. The use of jerk parameter is
equivalent to the statefinder used in \cite{sahni,alam2}. We may
use the above expression (\ref{lumdis}) to probe the geometry of
the Universe \cite{chiba,rrcmk}. Note that the Taylor expansion of
$d_{\rm L}$ may break down at high $z$ and the actual behaviour of
$d_{\rm L}$ may not be represented by finite number of terms. It
is also straightforward to get
\begin{equation}
\label{hubble} H^2(z)=H^2_0[1+2(1+q_0)z+(1+2q_0+j_0)z^2-{1\over
3}(s_0+q_0j_0)z^3+O(z^4)],
\end{equation}
\begin{equation}
\label{dec} q(z)=q_0-(q_0+2q_0^2-j_0)z+\left(4q_0^3+4q_0^2+q_0-
2j_0-{s_0\over 2}-{7j_0q_0\over 2}\right)z^2+O(z^3).
\end{equation}

Now let us find out $q_0$, $j_0$ and $s_0$ from the SN Ia data
compiled by Riess etal., These parameters are determined by
minimizing
\begin{equation} \label{lrmin} \chi^2=\sum_i{[\mu_{\rm
obs}(z_i)-\mu(z_i)]^2\over \sigma^2_i},
\end{equation}
where $\sigma_i$ is the total uncertainty in the SN Ia observation
and the extinction-corrected distance modulus
$\mu(z)=5\log_{10}(d_{\rm L}(z)/{\rm Mpc})+25$. In the fitting
process, we use the SN Ia gold sample only and we consider a flat
universe with $\Omega_k=0$. Because we use Taylor expansion to get
the luminosity distance, this expansion may break down at high
$z$. Therefore, we first use the full 157 gold sample SNe, then we
use those 148 SNe with $z\le 1.0$. The best fit parameters to the
whole 157 gold sample SNe are ($q_0$, $j_0$, $s_0$)=($-1.1$,
$6.4$, $39.5)$ with $\chi^2=174.2$. The best fit parameters to the
148 gold sample SNe with $z\le 1.0$ are ($q_0$, $j_0$,
$s_0$)=($-1.7$, $14.4$, $149.4)$ with $\chi^2=160.8$.

\begin{figure}
\centering
\includegraphics[width=8cm]{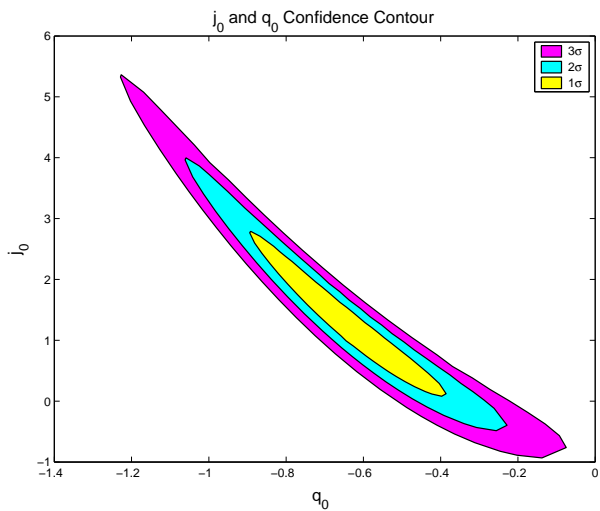}
\caption{The plot of $q_0$ and $j_0$ contour fitting to the whole
157 gold sample SNe.} \label{cont1}
\end{figure}

\begin{figure}
\centering
\includegraphics[width=8cm]{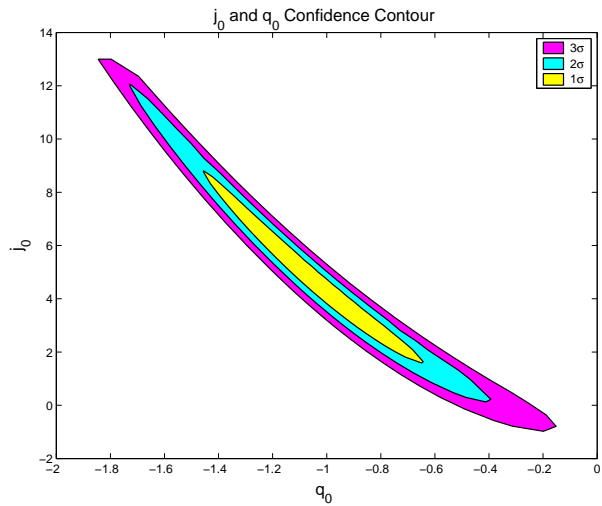}
\caption{The plot of $q_0$ and $j_0$ contour fitting to the 148
gold sample SNe with $z\le 1.0$.} \label{cont2}
\end{figure}

If we expand the luminosity distance $d_{\rm L}$ to the third
order only, i.e., we only consider the parameters $q_0$ and $j_0$
in Eq. (\ref{lumdis}), then we find that the best fit parameters
to the whole 157 gold sample SNe are: $q_0=-0.64^{+0.25}_{-0.26}$,
$j_0=1.2^{+1.5}_{-1.1}$ and $\chi^2=176.1$. At 99.5\% confidence
level, $q_0=-0.64^{+0.56}_{-0.59}$, so we conclude that the
expansion of the Universe is accelerating with 99.5\% confidence.
From Eq. (\ref{dec}), we get $z_{\rm
T}=q_0/(q_0+2q_0^2-j_0)=0.595_{-0.177}^{+1.849}$. The best fit
parameters to the 148 gold sample SNe with $z\le 1.0$ are:
$q_0=-1.0\pm 0.4$, $j_0=4.7^{+4.1}_{-3.1}$ and $\chi^2=161.3$. At
99.5\% confidence level, $q_0=-1.0^{+0.9}_{-1.0}$, so we conclude
again that the expansion of the Universe is accelerating with
99.5\% confidence. With the best fit parameters, we find that
$z_{\rm T}=q_0/(q_0+2q_0^2-j_0)=0.295_{-0.056}^{+0.174}$. The
contour plot for $q_0$ and $j_0$ is shown in Figs. \ref{cont1} and
\ref{cont2}.

So far our analysis uses the FRW metric only, we have not
specified any gravitational theory yet. The above results are
applicable to a wide range of theories. For example,
$q_0=(\Omega_{\rm m0}-2\Omega_\Lambda)/2$ and $j_0=\Omega_{\rm
m0}+\Omega_\Lambda$ for the $\Lambda$-CDM model. If we expand the
luminosity distance to the fifth order with the crackle parameter
$c(t)=(aH^5)^{-1}d^5a/dt^5$, then we need to add to Eq.
(\ref{lumdis}) the following correction
\begin{eqnarray}
\label{5order}
 -{1\over
120}(6+14q_0-61q^2_0-160q^3_0-105q^4_0+110q_0j_0\nonumber\\+105q^2_0j_0
+15q_0s_0+27j_0-10j^2_0+11s_0+c_0)z^5.
\end{eqnarray} The best fit
parameters to the whole 157 gold sample SNe are ($q_0$, $j_0$,
$s_0$, $c_0$)=($-1.1$, $7.6$, $55.6$, 676.6) with $\chi^2=173.4$.
The correction to $H_0d_{\rm L}$ at $z=1.5$ is about $-1.3$ which
is around 34\%. The best fit parameters to the 148 gold sample SNe
with $z\le 1.0$ are ($q_0$, $j_0$, $s_0$, $c_0$)=($-1.5$, $11.4$,
$101.2$,$1484.2$) with $\chi^2=160.8$. The correction to
$H_0d_{\rm L}$ at $z=1.5$ is about $0.19$ which is around 8.5\%.
Therefore the introduction of the fifth order correction changes
the value of $q_0$ a little. We still have $q_0<0$. It is clear
that the kinematic determination of the cosmological parameters is
better suited for low redshift SNe Ia. However, from the
observational data, $(z,\ \mu(z))=(1.4,\ 45.09)$, $(z,\
\mu(z))=(1.551,\ 45.3)$ and $(z,\ \mu(z))=(1.755,\ 45.53)$, we see
that $\Delta\mu(z)=0.21$ when $\Delta z=0.151$ and
$\Delta\mu(z)=0.23$ when $\Delta z=0.205$. Theoretically, we know
that $d\mu(z)=(5/\ln(10))(d\,d_{\rm L}(z)/d_{\rm L}(z)dz)dz$. From
Eq. (\ref{lumdis}), we get $d\mu(z)=(5/\ln(10))(n/z)dz$ if the
luminosity distance is dominated by the higher term $z^n$.
Combining the above analysis, we find that $n\sim 1$. Therefore,
in this case, the higher term may not be the dominant term.

We conclude that $q_0<0$ with 99.5\% confidence. In other words,
we conclude that the Universe is expanding with acceleration.

\section{"Taylor expansion" of Dark Energy Density}
\begin{figure}
\centering
\includegraphics[width=8cm]{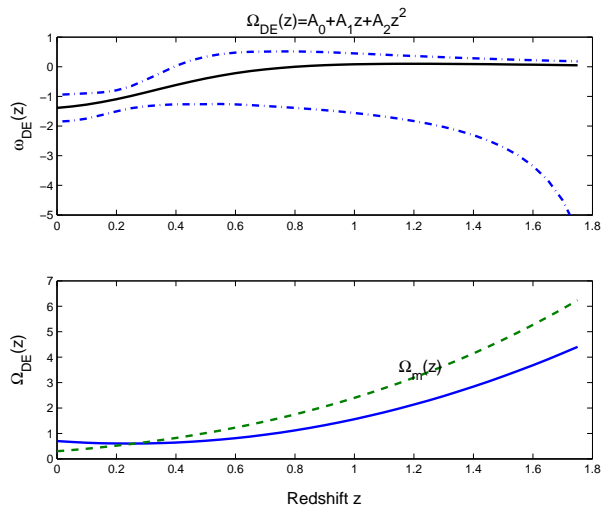}
\caption{The best fit to the 157 gold sample SNe Ia with the prior
$\Omega_{\rm m0}=0.3\pm 0.04$. The upper panel shows $\omega_{\rm
DE}(z)$, the dotted dash lines are the 1$\sigma$ regions. The
lower panel shows $\Omega_{\rm m}(z)$ and $\Omega_{\rm DE}(z)$}
\label{wqzp}
\end{figure}

In this section, we parameterize the dark energy density as
\cite{alam} \begin{equation} \label{poly2} \Omega_{\rm
DE}(z)=A_0+A_1(1+z)+A_2(1+z)^2,\end{equation} where $\Omega_{\rm
DE}(z)=8\pi G\rho_{\rm DE}(z)/(3H^2_0)$, $\Omega_{\rm m0}=8\pi
G\rho_{\rm m0}/(3H^2_0)$ and $A_0=1-\Omega_{\rm m0}-A_1-A_2$. This
parametrization is equivalent to Eq. (\ref{hubble}) with
$\Omega_{\rm m0}=-(s_0+q_0j_0)/3$. The relationship between
$\omega_{\rm DE}$ and $z$ is
$$\omega_{\rm DE}={1+z\over
3}{A_1+2A_2(1+z)\over A_0+A_1(1+z)+A_2(1+z)^2}-1.$$ With the above
parameteriaztion, we find that $\Omega_{\rm DE}\ll \Omega_{m}$ and
$\omega_{\rm DE}\approx -1/3$ when $z\gg 1$. Combining the above
two equations, we find that the transition redshift $z_{\rm T}$
satisfies \begin{equation} \label{ztpoly} \Omega_{\rm
m0}(1+z)^3-A_1(1+z)-2A_0=0. \end{equation} The best fit to the
whole 157 gold sample SNe gives $A_1=-5.79$, $A_2=2.9$ and
$\Omega_{\rm m0}\sim 0$ with $\chi^2=174.05$. If we use a Gaussian
prior $\Omega_{\rm m0}=0.3\pm 0.04$ \cite{tegmark}, then we get
the best fit parameters $A_1=-4.2^{+4.6}_{-5.4}$ and
$A_2=1.7^{+2.2}_{-1.8}$ with $\chi^2=174.21$. Substitute these
parameter values to Eq. (\ref{ztpoly}), we find that $z_{\rm
T}=0.35$. The evolutions of the dark energy density and
$\omega_{\rm DE}$ are shown in Fig. \ref{wqzp}. Alam et al. showed
that the SNe Ia data favored an evolving dark energy model by
using the above reconstruction \cite{alam,alam1}. They also showed
that $z_{\rm T}\sim 0.4$. Our results are consistent with those
analysis.

\begin{figure}
\centering
\includegraphics[width=8cm]{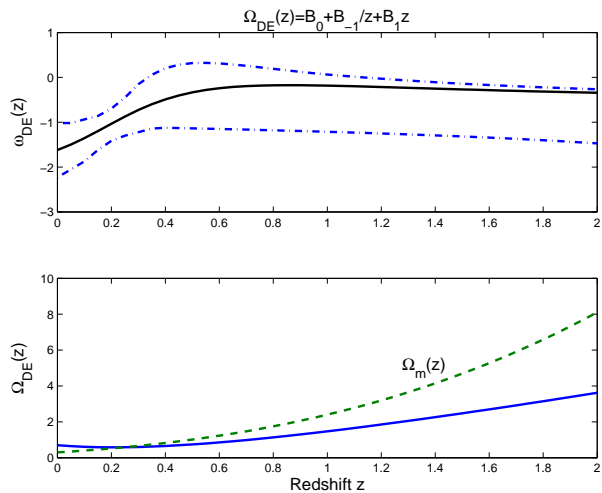}
\caption{The best fit to the 157 gold sample SNe Ia with the prior
$\Omega_{\rm m0}=0.3\pm 0.04$. The upper panel shows $\omega_{\rm
DE}(z)$, the dotted dash lines are the 1$\sigma$ regions. The
lower panel shows $\Omega_{\rm m}(z)$ and $\Omega_{\rm DE}(z)$}
\label{wqzpn}
\end{figure}

Because it is possible that $\omega_{\rm DE}<-1$, so we consider
another two parameter representation of dark energy
\begin{equation} \label{neg} \Omega_{\rm
DE}(z)=B_0+B_1(1+z)+B_{-1}/(1+z),\end{equation} where
$B_0=1-\Omega_{\rm m0}-B_1-B_{-1}$. with this parametrization, we
get
$$\omega_{\rm DE}={1\over
3}{B_1(1+z)^2-B_{-1}\over B_1(1+z)^2+B_0(1+z)+B_{-1}}-1.$$ The
above equation tells us that $\Omega_{\rm DE}\ll \Omega_{m}$ and
$\omega_{\rm DE}\approx -2/3$ when $z\gg 1$. Combining the above
two equations, we find that the transition redshift $z_{\rm T}$
satisfies \begin{equation} \label{ztneg} \Omega_{\rm
m0}(1+z)^3-B_1(1+z)-2B_0-{3B_{-1}\over 1+z}=0. \end{equation} The
best fit to the whole 157 gold sample SNe Ia gives $B_{-1}=6.87$,
$B_1=6.14$ and $\Omega_{\rm m0}\sim 0$ with $\chi^2=173.2$. If we
use a Gaussian prior $\Omega_{\rm m0}=0.3\pm 0.04$, then we get
the best fit parameters $B_{-1}=4.1^{+4.7}_{-4.2}$ and
$B_1=2.8^{+3.4}_{-3.0}$ with $\chi^2=173.65$. Substitute the best
fit parameters into Eq. (\ref{ztneg}), we get $z_{\rm T}=0.30$.
The evolutions of $\omega_{\rm DE}$ and $\Omega_{\rm DE}$ are
shown in Fig. \ref{wqzpn}.

\section{Stable Parametrization}
\begin{figure}
\centering
\includegraphics[width=8cm]{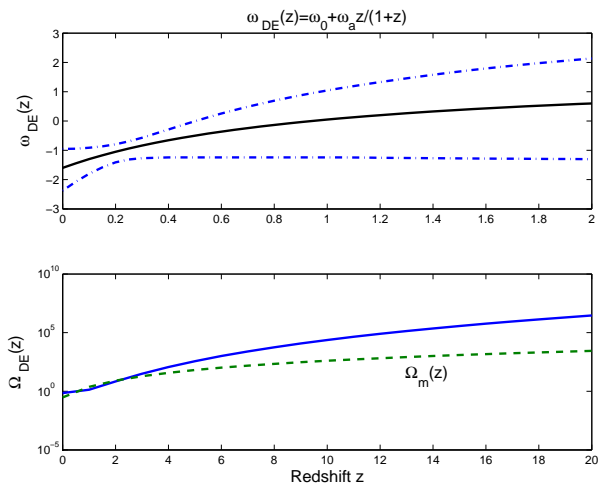}
\caption{The best fit to the 157 gold sample SNe Ia with the prior
$\Omega_{\rm m0}=0.3\pm 0.04$. The upper panel shows $\omega_{\rm
DE}(z)$, the dotted dash lines are the 1$\sigma$ regions. The
lower panel shows $\Omega_{\rm m}(z)$ and $\Omega_{\rm DE}(z)$}
\label{wqzl}
\end{figure}

In this section, we first consider the parametrization
\cite{polarski,linder} \begin{equation}\label{linder}\omega_{\rm
DE}=\omega_0+{\omega_a z\over 1+z}.\end{equation} When $z\gg 1$,
we have $\omega_{\rm DE}\sim \omega_0+\omega_a$. The dark energy
density is
$$\Omega_{\rm DE}=\Omega_{\rm
DE0}(1+z)^{3(1+\omega_0+\omega_a)}\exp(-3\omega_az/(1+z)).$$
Combining the above two equations, we find that $z_{\rm T}$
satisfies
\begin{equation}
\label{ztlind}
\Omega_{\rm
m0}+(1-\Omega_{\rm m0})\left(1+3\omega_0+{3\omega_a z\over
1+z}\right)(1+z)^{3(\omega_0+\omega_a)}
\exp\left({-3\omega_az\over 1+z}\right)=0.
\end{equation}
The best fit to the whole 157 gold sample SNe Ia gives
$\omega_0=-2.5$, $\omega_a=3.7$ and $\Omega_{\rm m0}=0.46$ with
$\chi^2=173.5$. If we use a Gaussian prior $\Omega_{\rm m0}=0.3\pm
0.04$, then we get the best fit parameters
$\omega_0=-1.6^{+0.6}_{-0.8}$ and $\omega_a=3.3^{+3.4}_{-3.7}$
with $\chi^2=173.92$. Substitute the best fit parameters into Eq.
(\ref{ztlind}), we get $z_{\rm T}=0.35$. The evolutions of
$\omega_{\rm DE}$ and $\Omega_{\rm DE}$ are shown in Fig.
\ref{wqzl}.

\begin{figure}
\centering
\includegraphics[width=8cm]{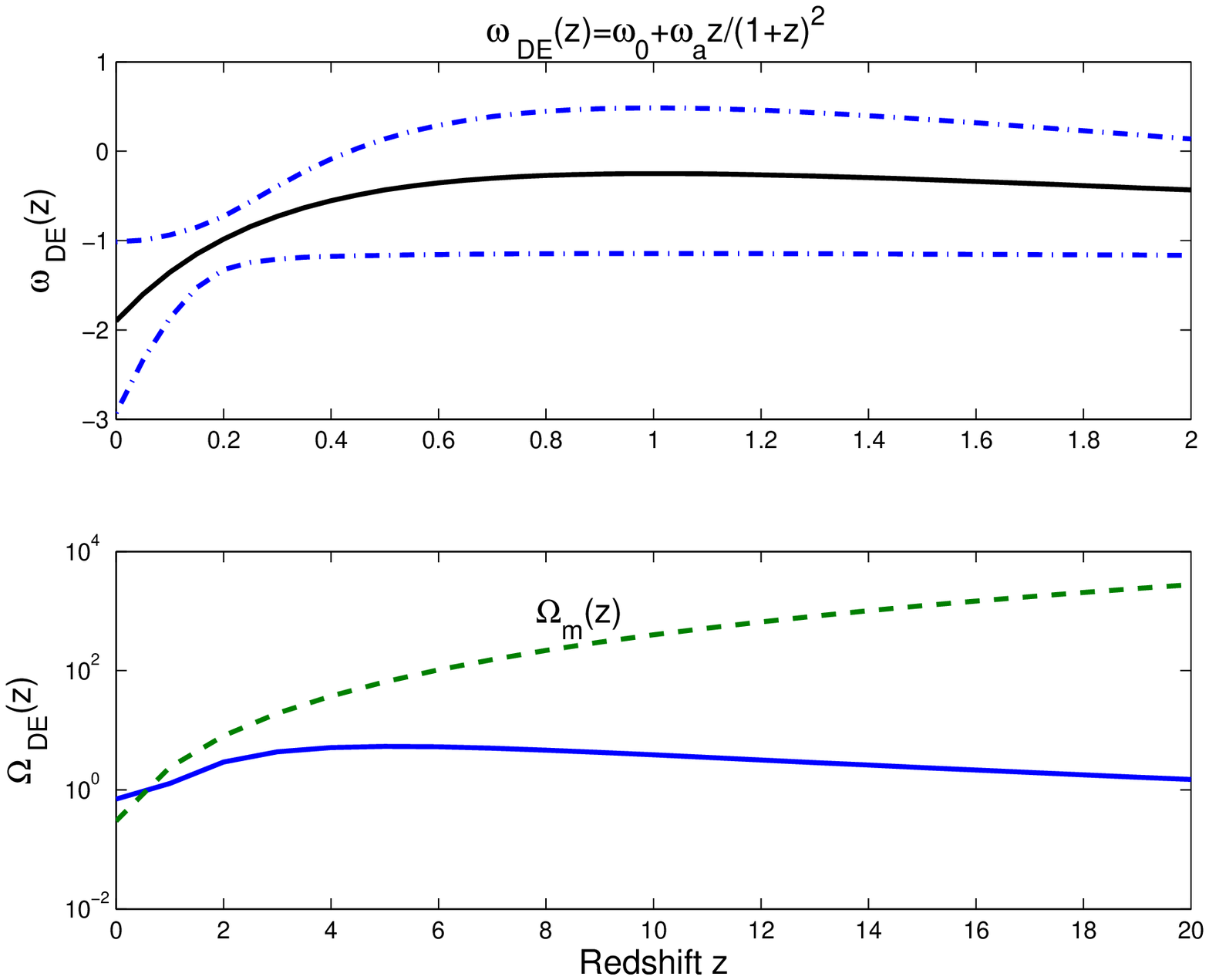}
\caption{The best fit to the 157 gold sample SNe Ia with the prior
$\Omega_{\rm m0}=0.3\pm 0.04$. The upper panel shows $\omega_{\rm
DE}(z)$, the dotted dash lines are the 1$\sigma$ regions. The
lower panel shows $\Omega_{\rm m}(z)$ and $\Omega_{\rm DE}(z)$}
\label{wqzl2}
\end{figure}

From Fig. \ref{wqzl}, we see that the dark energy density is
greater than the matter density at high $z$ because
$\omega_0+\omega_a>0$. So this stable parametrization may not be a
good choice at high $z$. Recently, Jassal, Bagla and Padmanabhan
considered the following parametrization \cite{hkjbp},
\begin{equation}\label{jbp}\omega_{\rm DE}=\omega_0+{\omega_a
z\over (1+z)^2}.\end{equation} When $z\gg 1$, we have $\omega_{\rm
DE}\sim \omega_0$. The dark energy density is
$$\Omega_{\rm DE}=\Omega_{\rm
DE0}(1+z)^{3(1+\omega_0)}\exp(3\omega_az^2/2(1+z)^2).$$
Combining
the above two equations, we find that $z_{\rm T}$ satisfies
\begin{equation}
\label{ztjbp} \Omega_{\rm m0}+(1-\Omega_{\rm
m0})\left(1+3\omega_0+{3\omega_a z\over (1+z)^2}\right)
(1+z)^{3\omega_0}\exp\left({3\omega_az^2\over 2(1+z)^2}\right)=0.
\end{equation}
The best fit to the whole 157 gold sample SNe Ia gives
$\omega_0=-2.5$, $\omega_a=7.6$ and $\Omega_{\rm m0}=0.42$ with
$\chi^2=173.3$. If we use a Gaussian prior $\Omega_{\rm m0}=0.3\pm
0.04$, then we get the best fit parameters
$\omega_0=-1.9^{+0.9}_{-1.1}$ and $\omega_a=6.6\pm 6.7$ with
$\chi^2=173.41$. Substitute the best fit parameters into Eq.
(\ref{ztjbp}), we get $z_{\rm T}=0.30$. The evolutions of
$\omega_{\rm DE}$ and $\Omega_{\rm DE}$ are shown in Fig.
\ref{wqzl2}.
From Fig. \ref{wqzl2}, it is clear that the dark energy density
did not dominate over the matter energy density at high $z$. Our
result is consistent with that obtained in \cite{hkjbp}.
\section{Wetterich's Parametrization}
\begin{figure}
\centering
\includegraphics[width=8cm]{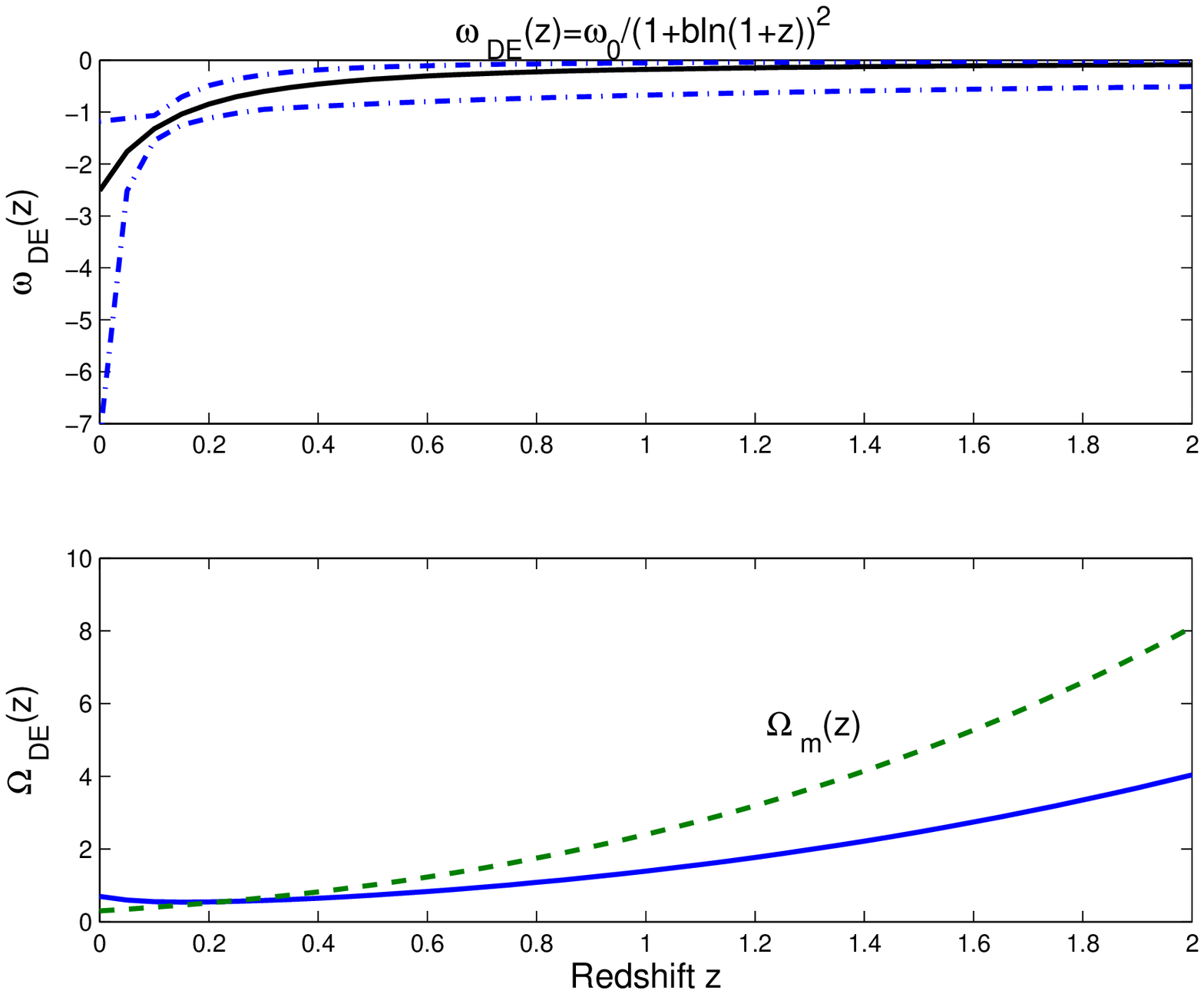}
\caption{The best fit to the 157 gold sample SNe Ia with the prior
$\Omega_{\rm m0}=0.3\pm 0.04$. The upper panel shows $\omega_{\rm
DE}(z)$, the dotted dash lines are the 1$\sigma$ regions. The
lower panel shows $\Omega_{\rm m}(z)$ and $\Omega_{\rm DE}(z)$}
\label{wqzw}
\end{figure}

\begin{figure}
\centering
\includegraphics[width=8cm]{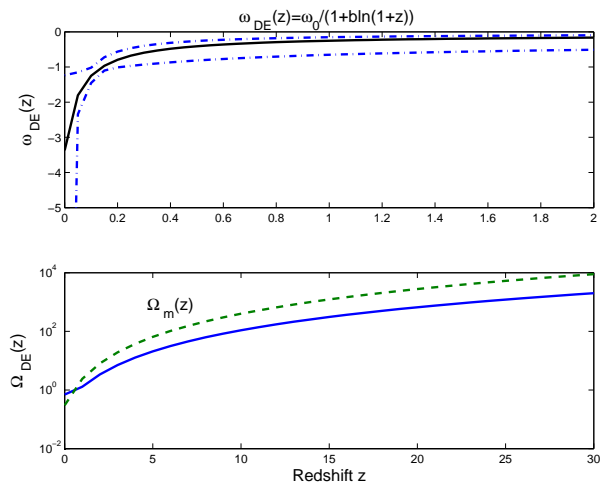}
\caption{The best fit to the 157 gold sample SNe Ia with the prior
$\Omega_{\rm m0}=0.3\pm 0.04$. The upper panel shows $\omega_{\rm
DE}(z)$, the dotted dash lines are the 1$\sigma$ regions. The
lower panel shows $\Omega_{\rm m}(z)$ and $\Omega_{\rm DE}(z)$}
\label{wqzna}
\end{figure}

In this section, we first consider the parametrization given in
\cite{wetterich}, \begin{equation}\label{wet}\omega_{\rm
DE}={\omega_0\over [1+b\ln(1+z)]^2}.\end{equation} When $z\gg 1$,
we have $\omega_{\rm DE}\sim 0$. The dark energy density is
 $$\Omega_{\rm
DE}=\Omega_{\rm DE0}(1+z)^{3+3\omega_0/[1+b\ln(1+z)]}.$$
Combining
the above two equations, we find that $z_{\rm T}$ satisfies
\begin{equation}
\label{ztwet} \Omega_{\rm m0}+(1-\Omega_{\rm
m0})\left(1+{3\omega_0\over [1+b\ln(1+z)]^2}\right)
(1+z)^{3\omega_0/[1+b\ln(1+z)]}=0.
\end{equation}
The best fit to the whole 157 gold sample SNe Ia gives
$\omega_0=-1.84$, $b=5.85$ and $\Omega_{\rm m0}\sim 0$ with
$\chi^2=173.09$. If we use a Gaussian prior $\Omega_{\rm
m0}=0.3\pm 0.04$, then we get the best fit parameters
$\omega_0=-2.5^{+1.3}_{-4.8}$ and $b=4.0^{+11.4}_{-3.5}$ with
$\chi^2=173.15$. Substitute the best fit parameters into Eq.
(\ref{ztwet}), we get $z_{\rm T}=0.26$. The evolutions of
$\omega_{\rm DE}$ and $\Omega_{\rm DE}$ are shown in Fig.
\ref{wqzw}.
Because the best fit of the above parametrization gives
$\Omega_{\rm m0}\sim 0$ which is not physical, we first modify the
above parametrization as \begin{equation} \label{wet1} \omega_{\rm
DE}={\omega_0\over 1+b\ln(1+z)}.\end{equation} When $z\gg 1$, we
have $\omega_{\rm DE}\sim 0$. The dark energy density is
$$\Omega_{\rm DE}=\Omega_{\rm
DE0}(1+z)^3[1+b\ln(1+z)]^{3\omega_0/b}.$$ Combining the above two
equations, we find that $z_{\rm T}$ satisfies
\begin{equation} \label{ztwet1} \Omega_{\rm m0}+(1-\Omega_{\rm
m0})\left(1+{3\omega_0\over 1+b\ln(1+z)}\right)
\left[1+b\ln(1+z)\right]^{3\omega_0/b}=0.
\end{equation}
The best fit to the whole 157 gold sample SNe Ia gives
$\omega_0=-3.05$, $b=36.8$ and $\Omega_{\rm m0}\sim 0$ with
$\chi^2=172.75$. If we use a Gaussian prior $\Omega_{\rm
m0}=0.3\pm 0.04$, then we get the best fit parameters
$\omega_0=-3.4^{+2.1}_{-17.7}$ and
$\omega_a=17.8^{+162.3}_{-16.4}$ with $\chi^2=172.91$. This
modification does not solve the problem of $\Omega_{\rm m0}\sim
0$. Substitute the best fit parameters into Eq. (\ref{ztwet1}), we
get $z_{\rm T}=0.25$. The evolutions of $\omega_{\rm DE}$ and
$\Omega_{\rm DE}$ are shown in Fig. \ref{wqzna}.

\begin{figure}
\centering
\includegraphics[width=8cm]{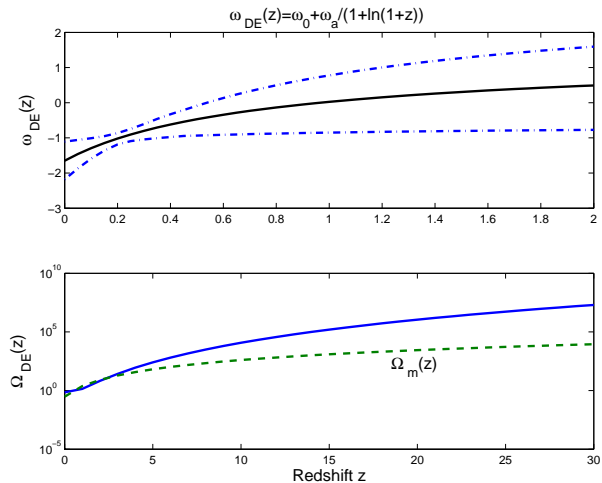}
\caption{The best fit to the 157 gold sample SNe Ia with the prior
$\Omega_{\rm m0}=0.3\pm 0.04$. The upper panel shows $\omega_{\rm
DE}(z)$, the dotted dash lines are the 1$\sigma$ regions. The
lower panel shows $\Omega_{\rm m}(z)$ and $\Omega_{\rm DE}(z)$}
\label{wqzn}
\end{figure}

Now let us consider another modification \begin{equation}
\label{wet2} \omega_{\rm DE}=\omega_0+{\omega_a\over
1+\ln(1+z)}.\end{equation} When $z\gg 1$, we have $\omega_{\rm
DE}\sim \omega_0$. The dark energy density is
$$\Omega_{\rm DE}=\Omega_{\rm
DE0}(1+z)^{3(1+\omega_0)}[1+\ln(1+z)]^{3\omega_a}.$$  Combining
the above two equations, we find that $z_{\rm T}$ satisfies
\begin{equation} \label{ztnew}
\Omega_{\rm m0}+(1-\Omega_{\rm
m0})\left(1+3\omega_0+{3\omega_a\over 1+b\ln(1+z)}\right)
(1+z)^{3\omega_0}[1+\ln(1+z)]^{3\omega_a}=0.
\end{equation}
The best fit to the whole 157 gold sample SNe Ia gives
$\omega_0=2.2$, $\omega_a=-4.7$ and $\Omega_{\rm m0}=0.454$ with
$\chi^2=173.47$. If we use a Gaussian prior $\Omega_{\rm
m0}=0.3\pm 0.04$, then we get the best fit parameters
$\omega_0=2.4^{+2.6}_{-2.9}$ and $\omega_a=-4.1^{+3.4}_{-3.1}$
with $\chi^2=173.81$. Substitute the best fit parameters into Eq.
(\ref{ztnew}), we get $z_{\rm T}=0.34$. The evolutions of
$\omega_{\rm DE}$ and $\Omega_{\rm DE}$ are shown in Fig.
\ref{wqzn}.
Although this modification solves the problem of $\Omega_{\rm
m0}\sim 0$, it is not good at early times because the dark energy
density dominated over the matter energy density at early times as
shown in Fig. \ref{wqzn}.
\section{Discussions}
\begin{figure}
\centering
\includegraphics[width=8cm]{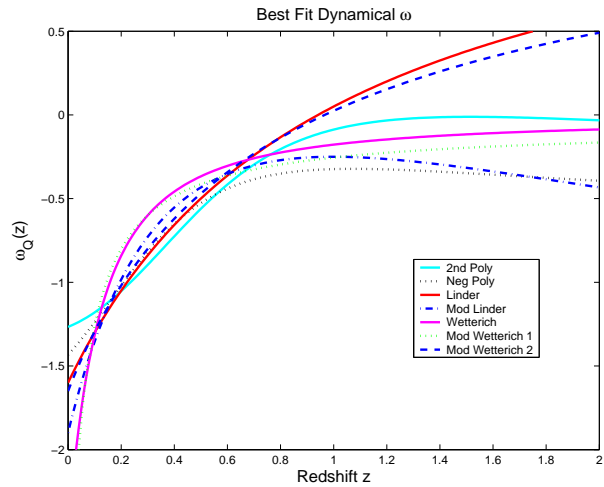}
\caption{The evolution of $\omega_{\rm DE}$ for different
parametrizations.} \label{rswqz}
\end{figure}

The SN Ia data shows that the expansion of the Universe is
accelerating. This conclusion derived from Eqs. (\ref{lumdis}) and
(\ref{5order}) does not dependent on any particular model. We used
the parametrizations (\ref{poly2}), (\ref{neg}), (\ref{linder}),
(\ref{jbp}) and (\ref{wet}) proposed in the literature to discuss
the property of dark energy. We also proposed two modified
parametrizations (\ref{wet1}) and (\ref{wet2}). By using the above
parametrizations, we derived the equations satisfied by the
transition redshift. In order to see the property of $\omega_{\rm
DE}(z)$, we re-plot $\omega_{\rm DE}(z)$ for all the models
considered in this paper together in Fig. \ref{rswqz}.
From Fig. \ref{rswqz}, we see that: (a) $\omega_{\rm DE0}<-1$.
This is also true at $1\sigma$ level. So the current SN Ia data
seems to marginally favor the dark energy metamorphosis suggested
in \cite{alam,alam1}. This does not mean that we can exclude the
$\Lambda$-CDM model; (b) $\omega_{\rm DE}(z)$ increases when $z$
increases. $\omega_{\rm DE}(z)$ changes more rapidly at low $z$
than at high $z$. This property may be due to the choice of the
parametrizations we made; (c) $z_{\rm T}\sim 0.3$. We also see
that the parametrization (\ref{jbp}) is a good choice. It avoids
the problem that the dark energy dominated the matter energy at
early times and the best fit $\Omega_{m0}$ to the SN Ia data for
this parametrization is not close to zero. The problem of
$\Omega_{m0}\sim 0$ is not a serious problem because $\chi^2$
depends on $\Omega_{m0}$ weakly for all the models discussed in
this paper. Daly and Djorgovski found that $z_{\rm T}\sim 0.4$ by
using a model independent analysis \cite{daly,daly1}. In our
analysis, we used Friedmann equation and some priors to interpret
the SN Ia data. As shown in \cite{mersini}, the interpretation of
the observational data changes drastically if the priors are
removed. We would like to stress that the results obtained in this
paper are consistent with other model independent analyses
obtained in the literature
\cite{hkjbp,alam,alam1,daly,daly1,gong,feng,wang}.

\ack The author is grateful to V. Johri, V. Sahni and R.A. Daly
for fruitful comments. The author thanks the hospitality of the
Interdisciplinary Center for Theoretical Study at the University
of Science and Technology of China where part of this work was
discussed. The author is grateful to J.X. Lu, B. Wang, X.M Zhang
and X.J. Wang for helpful discussions. The work is supported by
CQUPT under grant Nos. A2003-54 and A2004-05, NNSFC under grant
No. 10447008 and CSTC under grant No. 2004BB8601.

\end{document}